\documentclass[prl,twocolumn,showpacs]{revtex4}
\usepackage{graphicx}
\usepackage{amsmath}
\usepackage{bm}
\usepackage[T1]{fontenc}
\date{\today}
\begin{document}
\title{Nitrides as spintronic materials} 
\
\author{Tomasz Dietl\footnote{Address in 2003: 
Institute of Experimental and 
Applied Physics, Regensburg University; supported by Alexander von Humboldt 
Foundation}}
\affiliation{Institute of Physics, Polish Academy of
Sciences, al.~Lotnik\'ow 32/46, 02-668 Warszawa, Poland}

\begin{abstract}
A report of progress in spintronics-related works involving group III nitrides is given emphasizing contradictory opinions concerning basics characteristics of these materials. The actual position of magnetic impurities in the GaN lattice as well as a possible role of magnetic precipitates is discussed.  The question whether the hole introduce by the Mn impurities is localized tightly on the Mn d-levels or rather on the hybridized p-d bonding states is addressed. The nature of spin-spin interactions and magnetic phases, as provided by theoretical and experimental findings, is outlined, and possible origins of the high temperature ferromagnetism observed in (Ga,Mn)N are presented. Experimental studies aiming at evaluating characteristic times of spin coherence and dephasing in GaN are described.
\end{abstract}
\maketitle

\section{Introduction}

Semiconductor spintronics aims at developing material systems, in which novel mechanisms of control over magnetization in magnetic compounds or over individual spins in semiconductor nanostructures could lead to new functionalities in classical and quantum information hardware, respectively. Today's spintronic research involves virtually all material families. However, particularly interesting appear to be ferromagnetic semiconductors, which combine complementary functionalities of ferromagnetic and semiconductor material systems. For instance, it can be expected that powerful methods developed to control the carrier concentration and spin polarization in semiconductor quantum structures could serve to tailor the magnitude and orientation of magnetization produced by the spins localized on the magnetic ions. This paper intends to supplement previous surveys on spintronics \cite{Spin} and ferromagnetic semiconductors \cite{Diet02} by presenting the current status of spintronics-related works involving group III nitrides. In particular, we will specify some unique properties of these compounds, which make them promising materials in the this context.  We will then discuss results of relevant theoretical and experimental studies. It will be emphasized that the field is highly controversial, even in the case of the archetypical compound (Ga,Mn)N. In fact, divergent models reviewed in this paper concern such basic characteristics of this compound as: (i) the nature of electronic states associated with the presence of the magnetic constituent; (ii) character of the dominant spin-dependent interaction between the Mn spins; (iii) the origin of high-temperature ferromagnetic phase observed in this and related systems. We will also briefly describe experimental studies aiming at evaluating characteristic times of spin coherence and dephasing in GaN. On the other hand, in a sense spintronic light emitters, involving f-shells of rare earth elements, will not be discussed here.

\section{Why nitride DMS?}

There are specific properties of the nitride-based diluted magnetic semiconductors (DMS), which can result in a particularly high value of the Curie temperature. First, the small lattice constant leads to large hybridization between the valence orbitals and the d shells of the magnetic ions. Hence, the spin dependent interaction between the valence band holes and the localized spins, which controls the strength of the carrier-mediated ferromagnetic coupling, can be especially large in the nitrides. Secondly, a destructive influence of the spin-orbit interaction on this ferromagnetism is reduced significantly in materials containing light anions. These considerations, supported by theoretical modelling within the mean-field Zener model, led to the prediction that the Curie temperature in the Mn-based nitrides should exceeds 300~K, provided that it will be possible to develop materials containing sufficiently high concentrations of both valence band holes and substitutional Mn ions in the d$^5$ high spin configuration \cite{Diet00}, as shown in Fig.~1. The question of charge and spin states of magnetic impurities in the nitrides will be addressed in the next section. By contrast, we will not discuss the issue of p-type doping as certainly a number of papers in this volume addresses this question in detail. We remark only that the hole concentration can be enhanced by interfacial pyroelectric fields. Furthermore, there is some additional flexibility in pushing up solubility limits of both dopants and magnetic elements, as the epitaxial growth of cubic and hexagonal compounds involves, in general, different crystallographic surfaces. 

\begin{figure}
\includegraphics[width=90mm]{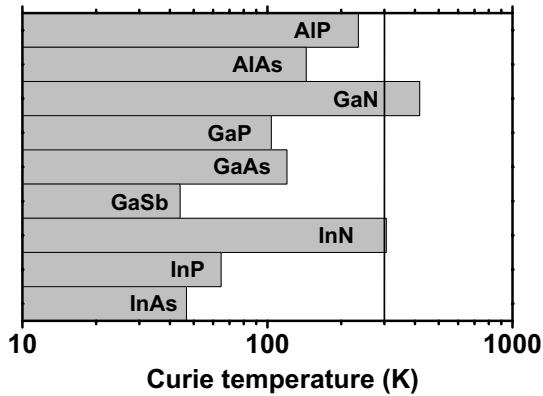}
\caption[]{Computed values of the Curie temperature for various p-type III-V compounds containing 5\% of Mn in the $S=5/2$ high spin state and $3.5\times 10^{20}$ holes per cm$^{3}$ (after \cite{Diet00,Diet01}).}
\label{fig:TC_III_V}
\end{figure}

\section{Electronic states of transition metal impurities; GaN:Mn}

According to the internal reference rule \cite{Lang88,Zung86}, the positions of states derived from magnetic shells do not vary across the entire family of the III-V compounds if the valence band offsets between different compounds are taken into account. In Fig.~2, adapted from Ref.~\cite{Blin02}, the data for III-V DMS containing various transition metals are collected. The symbols D(0/+) and A(0/$-$) denote the donor and acceptor states, which correspond to the transformation of the {\em triply} ionized magnetic ions M$^{3+}$ into M$^{4+}$ and into M$^{2+}$ ions, respectively.  Obviously, the internal reference rule may serve only for the illustration of chemical trends and not for extracting the precise values of the ionization energies. Moreover, if the acceptor state lies below the top of the valence band, the ground state corresponds to a hydrogenic-like acceptor d$^{N+1}$+h, not to the d$^N$ state. Importantly, band carriers introduce by such magnetic ions can mediate exchange interactions between the parent spins. Obviously, energies of hydrogenic-like states follow the band edges, and by no means are described by the internal reference rule. An example of an experimental study aiming at characterizing Mn states in GaN is shown in Fig.~3.

\begin{figure}
\includegraphics[width=90mm]{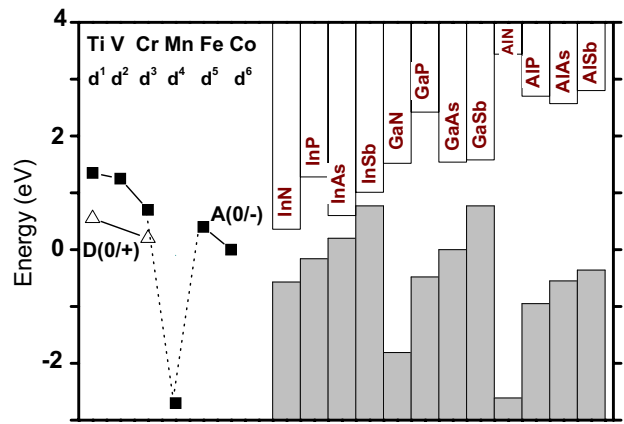}
\caption[]{Approximate positions of transition metals levels relative to the conduction band and valence band edges in various III-V compounds (adapted from \cite{Blin02}).}
\label{fig:TM}
\end{figure}

\begin{figure}
\includegraphics[width=85mm]{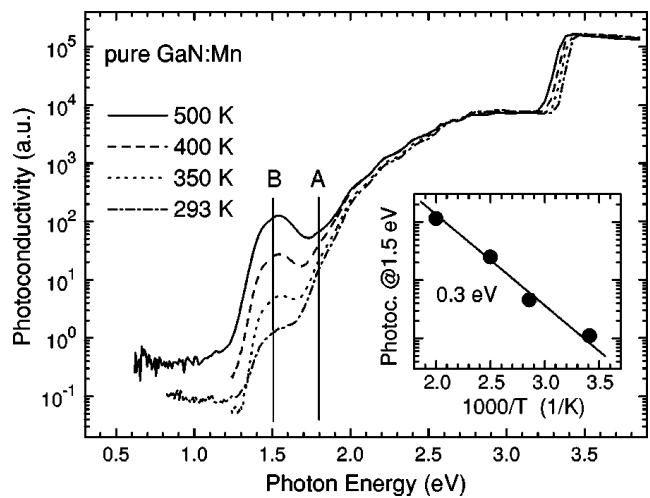}
\caption[]{Temperature dependence of photoconductivity spectra of GaN:Mn. The absorption, whose onset is marked by A, is interpreted as photoioniztion of the Mn acceptor-like state. The temperature dependent peak B is interpreted as acceptor photoionization process involving internal transitions within the Mn ions and thermally activated process characterized by an activation energy $E_A \approx 0.3$~eV (after Graf {\it et al.} \cite{Graf02}).}
\label{fig:Graf}
\end{figure}

Figure 4 summarizes ionization energies of the Mn acceptor level (Mn$^{3+}$/Mn$^{2+}$) across the III-V family, as evaluated by various authors from measurements of optical spectra and activation energy of conductivity. According to the discussion above, the Mn atom, when substituting a trivalent metal, may assume either of two configurations: (i) d$^4$ or (ii) d$^5$ plus a bound hole, d$^5$+h. The experimentally determined ionization energies correspond, therefore, to either d$^4$/d$^5$ or d$^5$+h/d$^5$ levels. It appears now to be a consensus that the Mn d states (t$_2$ and e$_g$ symmetry) reside in the valence band of antimonides and arsenides. Thus, in these materials the substitutional Mn  acts as an effective mass acceptor (d$^5$+h). Such a view is supported by electron spin resonance, photoemission, optical studies as well as by relatively small Mn concentrations leading to the insulator-to-metal transition \cite{Diet02}. The ground state has the t$_2$ symmetry, and its energy is determined by both Coulomb potential and "chemical shift". The latter results primarily from hybridization between Mn and neighboring anion states, which accounts for the valence band offset at high Mn concentrations.

\begin{figure}
\includegraphics[width=90mm]{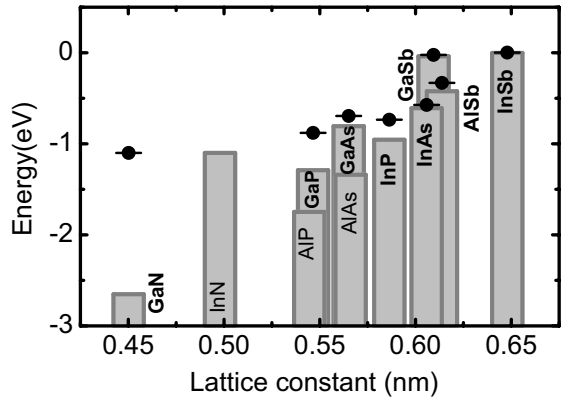}
\caption[]{Approximate positions of the acceptor Mn$^{2+}$/Mn$^{3+}$ levels relative to the valence band edge in various III-V compounds (adapted from \cite{Diet02a}).}
\label{fig:Mn}
\end{figure}

In contrast to antimonides and arsenides, the situation is much more intricate in the case of phosphides and nitrides. There is a growing amount of optical data indicating that the Mn gives rise to the presence of a mid-gap electron trap in GaN \cite{Koro01,Graf02}, as shown in Fig.~3. However, it is still uncertain whether the relevant state has d$^4$ ($S = 2$) or d$^5$+h ($J=1$) character. In particular, ESR measurements on GaN:Mn reveal only a line specific to d$^5$ ($S = 5/2$) Mn centers \cite{Graf02,Bara96,Zaja01} which are always present owing to a relatively high concentration of compensating donors. Even if the compensation is not overwhelming, the absence of other lines can be explained within either of the models, as both d$^4$ and d$^5$+h states contain an admixture of orbital momentum, which make the Zeeman splittings sensitive to local strains. Furthermore, a much greater optical cross section is observed \cite{Koro01,Graf02} for intra Mn$^{3+}$ excitations comparing to the case of Mn$^{2+}$ charge state. This is actually expected within both scenarios since in contrast to the d$^5$ case, the transitions within d$^4$ or d$^5$+h multiplets becomes either spin or parity allowed. On the other hand, the d$^4$ model appears to be supported by the {\em ab initio} computations within the local spin density approximation (LSDA), which points to the presence of the d-states in the gap of (Ga,Mn)N \cite{Gers01,Sato01,Schi01,Kula02,Kron02,Fili03}. However, the above interpretation has recently been call into question \cite{Diet02a}. In has been noted that a semi-empirical LSDA+U approach is necessary to reconcile the computed and photoemission positions of states derived from the Mn 3d shell in (Ga,Mn)As \cite{Okab01,Park00}. This implies that the Mn d states reside presumably in the valence band of GaN despite the 1.8~eV valence band offset between GaN and GaAs, as shown in Fig.~2. Within this scenario, a large hole binding energy results from the exceptionally strong spin-dependent p-d hybridization in the nitrides. Interestingly, the resulting charge transfer state has a character of a small magnetic polaron, and it is reminiscent of the Zhang-Rice singlet in cuprate superconductors.

\section{Spin-spin interactions and origin of ferromagnetism}

\begin{figure}
\includegraphics[width=80mm]{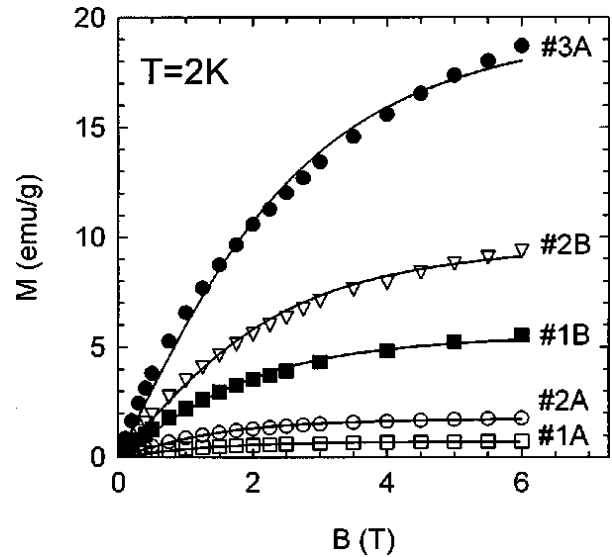}
\caption[]{Low temperature magnetization of Ga$_{1-x}$Mn$_x$N microcrystals (A) and bulk crystals (B)  with $0.0028 \le x \le 0.096$ showing paramagnetic behavior with week antiferromagnetic spin-spin interactions (after Zaj\c{a}c {\it et al.} \cite{Zaja01a}).}
\label{fig:Zajac}
\end{figure}

The status of experimental search for high $T_{\mbox{\small C}}$ ferromagnetic semiconductors has recently been reviewed \cite{Pear03}. Usually, epitaxy is employed but in Warsaw two methods for growth of bulk crystals \cite{Grze01,Szys01} and one yielding microcrystals \cite{Dwil98} are now being used in order to obtain (Ga,Mn)N and related materials. According to EXAFS spectra, the Mn occupies primarily substitutional position \cite{Soo01}. In general, magnetization of (Ga,Mn)N consists of paramagnetic and ferromagnetic components \cite{Kuwa01,Reed01,Over01,Sono02,Sasa02,Thal02,Baik03,Dhar03,Zaja03,Chit03}. Comprehensive studies of the former \cite{Zaja01a,Dhar03a}, such as those presented in Fig.~5, demonstrate the presence of {\em antiferromagnetic} spin-spin interactions, characterized by the nearest neighbor exchange integral $|J_{NN}| \approx 2$~K,  about one order of magnitude smaller than that in Mn-based II-VI DMS \cite{Diet94}.  The ferromagnetic component is not always present but in some samples it can involve up to about 20\% of the Mn spins, and can persist up to temperatures as high as 940~K, according to data presented in Fig.~6. It has been detected in both wurzite \cite{Kuwa01,Reed01,Over01,Sono02,Sasa02,Thal02,Baik03,Dhar03} and cubic \cite{Chit03} films deposited by MBE as well as in (Ga,Mn)N microcrystals grown by ammonothermal method \cite{Zaja03}.  Ferromagnetic ordering appears to result in the anomalous Hall effect \cite{Kim03} but not in the magnetic circular dichroism \cite{Ando03}. It should be emphasized at this point that the usefulness of ferromagnetic semiconductors requires the existence of a coupling between ferromagnetic and semiconducting properties, independently of the actual microscopic structure of the system. The room temperature ferromagnetic order has been reported also in the case of (Ga,Cr)N \cite{Hash02,Park02} and (Ga,Gd)N \cite{Tera02}, while (Ga,Fe)N shows indications of ferromagnetism below 100~K \cite{Akin00}.

\begin{figure}
\includegraphics[width=85mm]{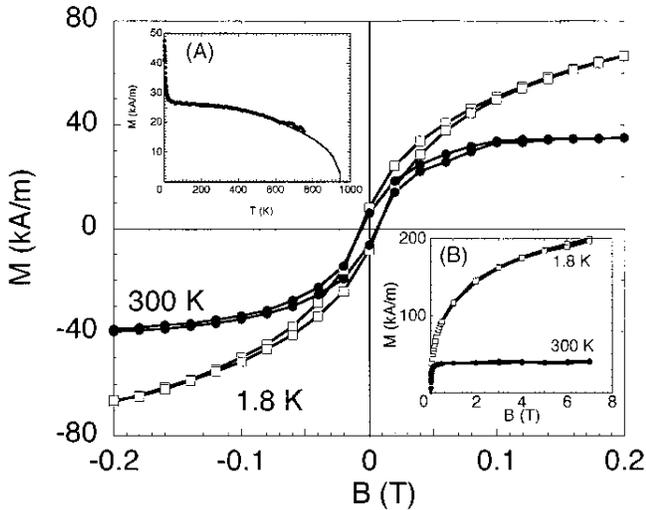}
\caption[]{Magnetization as a function of magnetic field and temperature (inset) of a (Ga,Mn)N film grown by MBE showing paramagnetic and ferromagnetic contributions. The solid line is calculated within the mean field approximation and points to the ferromagnetic transition temperature of 940 K (after Sasaki {\it et al.} \cite{Sasa02}).}
\label{fig:Sonoda}
\end{figure}

At present, the above findings can be interpreted in a number of ways. In particular, it is known that in the case of (Ga,Mn)As \cite{DeBo96,More02} and (Ga,Mn)Sb \cite{Abe00}, magnetization may contain a contribution from ferromagnetic precipitates (MnAs and MnSb, respectively), whose relative role and magnetic properties depend on the growth conditions. By the same token, ferrimagnetic Mn$_4$N \cite{Zaja03,Yang01} and related compounds are taken into consideration \cite{Rao02}. As shown in Fig.~7, the presence of some precipitates has indeed been detected in (Ga,Mn)N \cite{Dhar03}. Presumably, their chemical and magnetic properties will be soon determined, and it might become clear why the precipitates, despite their small size, exhibit ferromagnetic, not superparamagnetic behavior. 

\begin{figure}
\includegraphics[width=85mm]{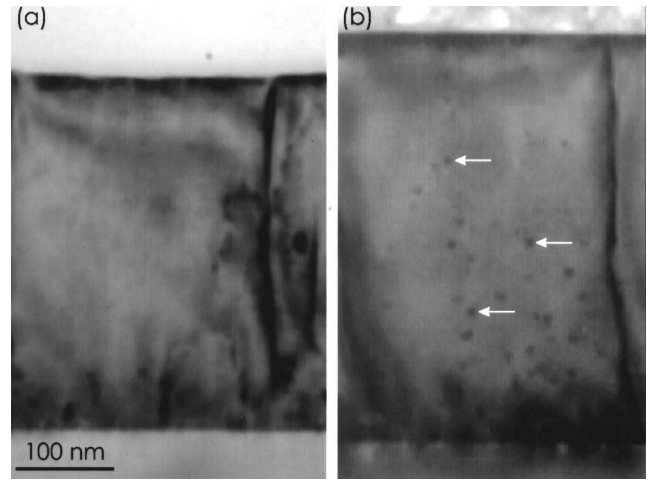}
\caption[]{Bright-field TEM micrograph of samples do not showing (a) and showing (b) a ferromagnetic behavior. The nm-scale clusters are visible in (b), some of which are highlighted by arrows (after Dhar {\it et al.} \cite{Dhar03}).}
\label{fig:Ploog}
\end{figure}

\begin{figure}
\includegraphics[width=95mm]{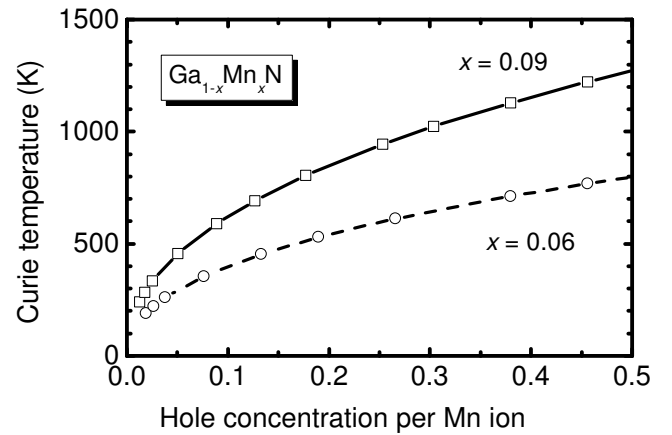}
\caption[]{The calculated Curie temperature of Ga$_{1-x}$Mn$_x$N as a function of the hole concentration in the valence band (after \cite{Diet02a}).}
\label{fig:TC}
\end{figure}

An interesting question then arises why the Ga$_{1-x}$Mn$_x$N alloy itself shows rather antiferromagnetic than ferromagnetic properties. Within the Zener model, the absence of ferromagnetism can be assigned to a strong localization of the holes. However, within {\em ab initio} LSDA theories, the ferromagnetism driven by the double exchange is expected even if the holes occupy the d$^4$ mid gap states \cite{Sato01,Schi01,Kula02,Kron02,Fili03}. Actually, it is possible that effects of the Mott-Anderson localization are underestimated within the LSDA and CPA and, accordingly, the role of the ferromagnetic double exchange is overestimated. Alternatively, a phenomenon of self compensation can be involved leading to the appearance of the Mn d$^5$ states, for which no ferromagnetism is to be expected. In fact, this was the case of first (In,Mn)As layers, which remained paramagnetic and n-type up Mn concentration $x = 0.18$ \cite{vonM91}. Similarly, it is now known that Mn interstitials act as double donors in (Ga,Mn)As, and reduce $T_{\mbox{\small C}}$ substantially in the regime of large Mn concentrations \cite{Yu02,Mase01,Blin03}. This leads us to still another scenario \cite{Diet02a}, developed within the d$^5$ + h model of the Mn gap states. If the compensation is not too strong, an overlap between the hole wave functions will result in the localization length of the holes greater than the distance between the Mn spins. Under such conditions, the Zener model becomes approximately valid, and can explain high values of $T_{\mbox{\small C}}$ without invoking precipitates of other compounds, as shown in Fig.~8 \cite{Diet02a,Jung02}. However, in order to explain only partial participation of the magnetic constituent in the ferromagnetic order, we suggest the existence of a phase segregation, which does not change tetrahedral coordination. Furthermore, a double exchange \cite{Sato01} or ferromagnetic superexchange \cite{Blin96} can be involved in some cases, particularly in compounds containing other magnetic elements than Mn.

 \section{Quantum spintronic devices}
 
It is known that weak spin decoherence and strong spin-spin coupling are required to fabricate functional quantum gates of semiconductor quantum dots. Owing to large energy gap and the weakness of spin-orbit interactions, especially long spin life times are to be expected in the nitrides. Figure 9 depicts results of time-resolved Faraday rotation, which has been used to measure electron spin coherence in n-type GaN epilayers \cite{Besc01}. Despite densities of charged threading dislocations of $5\times 10^8$ cm$^{-2}$, this coherence yields spin lifetimes of about 20~ns at temperatures of 5~K, and persists up to room temperature. Importantly, this low decoherence rate is coupled in GaN with a small value of the dielectric constant. This enhances characteristic energy scales for quantum dot charging as well as for the exchange interaction of the electrons residing on the neighboring dots. In view of persisting advances in nitride-based device processing, we may witness a rapid progress in this field in a near future.

\begin{figure}
\includegraphics[width=90mm]{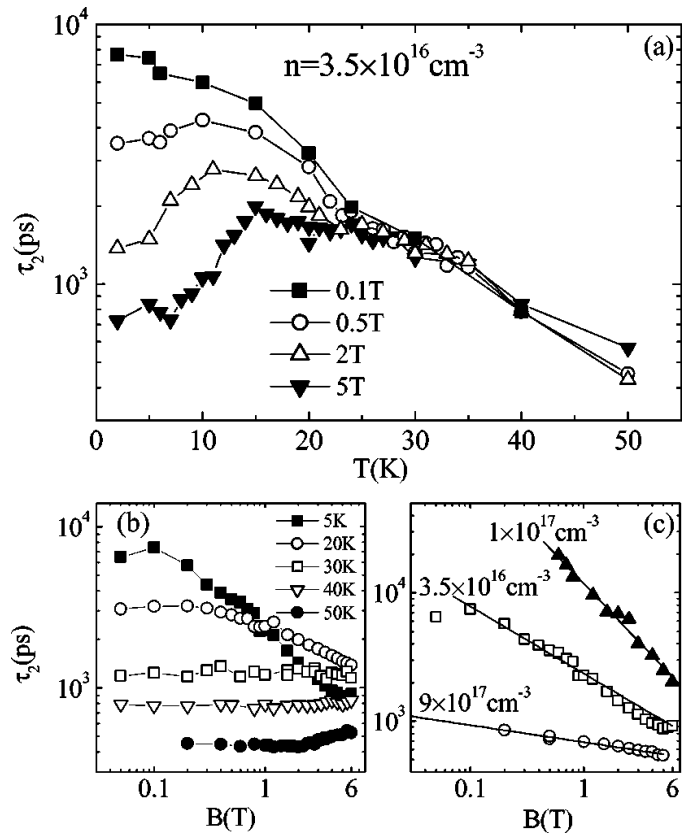}
\caption[]{Spin scattering time $\tau_2$ of n-GaN at various magnetic fields (a), temperatures (b)  ($n = 3.5\times 10^{16}$ cm$^{-3}$), and electron concentrations at 5 K (c) (after Beschoten {\it et al.} \cite{Besc01}).}
\label{fig:Beschoten}
\end{figure}

\section*{Acknowledgements}
The author would like to thank his co-workers, particularly F. Matsukura and H. Ohno in Sendai; J. Cibert and H. Mariette in Grenoble; P. Kacman, P. Kossacki, and M. Sawicki in Warsaw, A.H. MacDonald in Austin, and W. Wegscheider and D. Weiss in Regensburg for fruitful collaboration in studies of ferromagnetic semiconductors. Author's research in Germany in 2003 was supported by Alexander von Humboldt Foundation, while the work in Poland by State Committee for Scientific Research as well as by FENIKS project (EC G5RD-CT-2001-00535) within 5th Framework Programme of European Commission and Ohno Semiconductor Spintronics ERATO Project of Japan Science and Technology Corporation.


\begin{thebibliography}{10}

\bibitem{Spin} H. Ohno, F. Matsukura, Y. Ohno, JSAP International \textbf{5},  4 (2002) (available at http://www.jsapi.jsap.or.jp/); S. Wolf et al., Science \textbf{294}, 1488 (2001); T. Dietl, Acta Phys. Polon. A \textbf{100} (suppl.), 139 (2001) (available at http://xxx.lanl.gov/abs/cond-mat/0201279).
\bibitem{Diet02} T. Dietl, Semicond. Sci. Technol. \textbf{17}, 377 (2002); F. Matsukura, H. Ohno and T. Dietl, in: Handbook of Magnetic Materials, vol. 14, edited by K. H. J. Buschow (Elsevier, Amsterdam, 2002) p. 1.
\bibitem{Diet00} T. Dietl et al.,  Science \textbf{287}, 1019 (2000).
\bibitem{Diet01} T. Dietl, H. Ohno and F. Matsukura, Phys. Rev. B \textbf{63}, 195205 (2001).
\bibitem{Lang88} J. M. Langer et al., Phys. Rev. B \textbf{38}, 7723 (1988).
\bibitem{Zung86} A. Zunger in: {\it Solid State Phys.} vol. 39, edited by  H. Ehrenreich and  D. Turnbull  (New York, Academic Press, 1986) p. 275.
\bibitem{Blin02} J. Blinowski, P. Kacman and T. Dietl, in: Proceedings Material Research Society Symposia, vol. 690 (MRS, Pittsburg, 2002), p. F6.9.1; e-print: http://arXivorg/abs/cond-mat/0201012.
\bibitem{Koro01} R. Y. Korotkov et al., Physica B \textbf{308}, 18 (2001); {\it ibid} p. 30; Appl. Phys. Lett. \textbf{80}, 1731 (2002).
\bibitem{Graf02} T. Graf et al., Appl. Phys. Lett. \textbf{81}, 5159 (2002).
\bibitem{Diet02a} T. Dietl, F. Matsukura and H. Ohno, Phys. Rev. B \textbf{66}, 033203 (2002).
\bibitem{Bara96} P. G. Baranov et al., Semicond. Sci. Technol. \textbf{11}, 1843 (1996).
\bibitem{Zaja01} M. Zaj\c{a}c et al., Appl. Phys. Lett. \textbf{78}, 1276 (2001).
\bibitem{Gers01} U. Gerstmann, A. T. Blumenau and H. Overhof, Phys. Rev. B \textbf{63}, 075204 (2001).
\bibitem{Sato01} K. Sato and H. Katayama-Yoshida, Jpn. J. Appl. Phys. \textbf{40}, L485 (2001); Semicond. Sci. Technol. \textbf{17}, 367 (2002).
\bibitem{Schi01} M. van Schilfgaarde and O. N. Mryasov, Phys. Rev. B \textbf{63}, 233205 (2001).
\bibitem{Kula02} E. Kulatov et al., Phys. Rev. B \textbf{66}, 045203 (2002).
\bibitem{Kron02} L. Kronik, M. Jain and J. R. Chelikowsky, Phys. Rev. B \textbf{66}, 041203 (2002).
\bibitem{Fili03} A. Filippetti, N. A. Spaldin and S. Sanvito, e-print: http://arXivorg/abs/cond-mat/0302178. 
\bibitem{Okab01} J. Okabayashi et al., Phys. Rev. B \textbf{64}, 125304 (2001).
\bibitem{Park00} J. H. Park J H, S. K. Kwon and B. I. Min, Physica B \textbf{169}, 223 (2000).
\bibitem{Pear03} S. J. Pearton et al., J. Appl. Phys. \textbf{93}, 1 (2003).
\bibitem{Grze01} I. Grzegory, J. Phys.: Cond. Matt.  \textbf{13}, 6875 (2001).
\bibitem{Szys01} T. Szyszko et al., J. Crystal Growth \textbf{233}, 631 (2001).
\bibitem{Dwil98} R. Dwili\'nski et al., Diamond Relat. Mater. \textbf{7}, 1348 (1998).
\bibitem{Soo01} Y. L. Soo et al., Appl. Phys. Lett. \textbf{79}, 3926 (2001); M. Sato et al., Jpn. J. Appl. Phys. \textbf{41}, 4513 (2002).
\bibitem{Kuwa01} S. Kuwabara et al., Jpn. J. Appl. Phys. \textbf{40}, L724, 2001; Y. Shon et al., {\it ibid} 5304 (2001).
\bibitem{Theo01} N. Theodoropolpu et al., Appl. Phys. Lett. \textbf{78} 3475 (2001).
\bibitem{Reed01} M. L. Reed et al., Appl. Phys. Lett. \textbf{79}, 3473 (2001).
\bibitem{Over01} M. E. Overberg et al., Appl. Phys. Lett. \textbf{79}, 1312 (2001).
\bibitem{Sono02} S. Sonoda et al., J. Cryst. Growth \textbf{237-239},  1358 (2002).
\bibitem{Sasa02} T. Sasaki et al., J. Appl. Phys.  \textbf{91}, 7911 (2002).
\bibitem{Thal02} G. T. Thaler et al., Appl. Phys. Lett. \textbf{80}, 3964 (2002).
\bibitem{Baik03} J.M. Baik et al., Appl. Phys. Lett. \textbf{82}, 583 (2003).
\bibitem{Dhar03} S. Dhar et al., Appl. Phys. Lett. \textbf{82}, 2077 (2003).
\bibitem{Zaja03} M. Zaj\c{a}c et al., J. Appl. Phys.  \textbf{93}, 4715 (2003).
\bibitem{Chit03} V. Chitta et al., phys. stat. solidi, this Conference.
\bibitem{Hash02} M. Hashimoto et al., Solid State Commun. \textbf{122}, 37 (2002).
\bibitem{Park02} S. E. Park et al., Appl. Phys. Lett. (2002).
\bibitem{Tera02} N. Teraguchi et al., Solid State Commun. \textbf{122}, 651 (2002).
\bibitem{Akin00}  H. Akinaga et al., {\it ibid.} \textbf{77}, 4377 (2000)
\bibitem{Zaja01a} M. Zaj\c{a}c et al., Appl. Phys. Lett. \textbf{79}, 2432 (2001).
\bibitem{Dhar03a} S. Dhar et al., Phys. Rev. B \textbf{67} 165205 (2003).
\bibitem{Diet94} T. Dietl, in: Handbook on Semiconductors, vol. 3B ed. T. S. Moss (Elsevier, Amsterdam, 1994 ) p. 1251.
\bibitem{Kim03} K. H. Kim, Appl. Phys. Lett. \textbf{82}, 1775 (2003).
\bibitem{Ando03} K. Ando, Appl. Phys. Lett. \textbf{82}, 100 (2003).
\bibitem{DeBo96} J. De Boeck et al., Appl. Phys. Lett. \textbf{68}, 2744 (1996).
\bibitem{More02} M. Moreno et al., J. Appl. Phys. \textbf{92}, 4672 (2002).
\bibitem{Abe00} E. Abe et al., Physica E \textbf{7}, 981 (2000).
\bibitem{Rao02} B. K. Rao and P. Jena, Phys. Rev. Lett. \textbf{89}, 185504 (2002).
\bibitem{Yang01} H. Yang et al., Appl. Phys. Lett. \textbf{78}, 3860 (2001).
\bibitem{vonM91} S. von Moln\'ar et al., J. Magn. Magn. Mater. \textbf{93}, 356 (1991).
\bibitem{Yu02} K. M. Yu et al., Phys. Rev. B \textbf{65}, 201303(R) (2002).
\bibitem{Mase01} J. Masek and F. Maca, Acta Phys. Polon. A \textbf{100}, 319 (2001); F. Maca and J. Masek, Phys. Rev. B \textbf{65} 235209 (2002).
\bibitem{Blin03} J. Blinowski and P. Kacman, Phys. Rev. B \textbf{67}, 121204 (2003).
\bibitem{Jung02} T. Jungwirth et al., Phys. Rev. B \textbf{66}, 012402 (2002).
\bibitem{Blin96} J. Blinowski, P. Kacman and J. Majewski, Phys. Rev. B \textbf{53}, 9524 (1996); J. Cryst. Growth \textbf{159}, 972 (1996).
\bibitem{Besc01} B. Beschoten et al., Phys. Rev. B \textbf{63}, 121202 (2001).

\end{thebibliography}
\end{document}